\documentclass[preprint,showpacs,showkeys,preprintnumbers,amsmath,amssymb,dvipdfmx]{revtex4}

\usepackage[dvipdfmx]{graphicx}
\usepackage{dcolumn}
\usepackage{bm}
\usepackage[colorlinks]{hyperref}
\usepackage{slashbox}

\begin{document}

\title{More Tolerant Reconstructed Networks by Self-Healing 

against Attacks in Saving Resource
}

\author{Yukio Hayashi}
\affiliation{
Japan Advanced Institute of Science and Technology,\\
Ishikawa, 923-1292, Japan}
\author{Atsushi Tanaka}
\affiliation{Yamagata University\\
Yonezawa-city, Yamagata 992-8510, Japan}
\author{Jun Matsukubo}
\affiliation{National Institute of Technology Kitakyushu College\\
Kitakyushu-city, Fukuoka 802-0985, Japan
}

\date{\today}

\begin{abstract}
Complex network infrastructure systems for power-supply, 
communication, 
and transportation support our economical and social activities, 
however they are extremely vulnerable against the frequently increasing 
large disasters or attacks. Thus, a reconstructing from damaged network 
is rather advisable than empirically performed recovering to the original 
vulnerable one. In order to reconstruct a sustainable network, we focus 
on enhancing loops so as not to be trees as possible by node removals. 
Although this optimization is corresponded to an intractable combinatorial 
problem, we propose self-healing methods based on enhancing loops in 
applying an approximate calculation inspired from a statistical physics 
approach. We show that both higher robustness and efficiency are obtained 
in our proposed methods with saving the resource of links and ports than 
ones in the conventional healing methods. Moreover, the reconstructed 
network by healing can become more tolerant than the original one before
attacks, when some extent of damaged links are reusable or compensated as 
investment of resource. These results will be open up the potential of 
network reconstruction by self-healing with adaptive capacity in the 
meaning of resilience.
\end{abstract}

%

\maketitle



\section{Introduction}

Unfortunately, the frequencies of large disasters or malicious attacks 
increase due to climate exchange, crustal movements, 
military conflicts, cyber-terrorism, and 
mega-urbanization in our world day by day. 
For example, it is well known that 
a little accident involved the large area's power collapse 
in North America \cite{NERC04} or Italian peninsula \cite{UCTE} in 2003, 
and that enormous destruction of social infrastructure systems 
was happen by the great earthquake in Japan in 2011 \cite{WHO2013}.
While there exists a surprisingly common topological structure called 
scale-free (SF) in many real networks \cite{Amaral00}, 
such as power-grid, airline, 
communication, transportation systems, and so on, which support our 
social activities, economy, industrial production, etc. 
The SF structure is considered to be generated by a selfish rule: 
preferential attachment \cite{Barabasi99}, 
and consisted of many low degree nodes 
and a few (high degree) hubs, heterogeneously. 
Here, degree means the number of links at a node.
Moreover, by the heterogeneity, 
a SF network has extreme vulnerability against hub attacks \cite{Albert00}.
These vulnerable infrastructures appear everywhere and are 
interdependent on each other.
Exactly, since a node prefers to connect high degree nodes in the 
efficiency bias to shorten the path lengths counted by hops, 
the preferential attachment encourages the heavy concentration 
of links to hubs. 
In many real networks, 
once hubs are damaged and removed as malfunction, 
the remaining nodes are fragmented 
and lost the basic function for communication or transportation.
It is a plausible scenario for our network infrastructures 
that the weak points of hubs are involved in a large disaster.

Therefore, when large-scale failures or attacks occur, 
recovery to the original vulnerable network is inadvisable. 
Rather reconstruction by healing is required. 
In changing the structure instead of recovering to the original one, 
a question arises as to how a sustainable network should be reconstructed 
to maintain the network function. 
However, the resources of links (wire cables, wireless communication 
    or transportation lines between two nodes, etc.) 
and ports (channels or plug sockets at a node, etc.) 
are usually limited, the allocation should be controlled 
at the same time in the rewiring or additional investment for healing.
Such a reconstruction conforms with the concept of resilience in system 
engineering or ecology as a new supple approach 
to sustain basic objective and integrity 
even in encountering with the extreme change of situations or 
environments (e.g., by disasters or malicious attacks) 
for technological system, organization, or individual 
\cite{Hollnagel06}\cite{Folke06}\cite{Zolli12}. 

In this paper, through numerical simulation, 
we study how to reconstruct a sustainable network 
under limited resource, and propose effective self-healing methods 
based on enhancing loops through a local process around damaged parts.
In addition, 
we show the significant improvement form the previous study \cite{Hayashi20} 
to reduce the additional ports prepared 
in advance besides reusable ports.
The motivations for enhancing loops are as follows.
In percolation analysis, as a part of network science, 
it has been found that onion-like structure 
with positive degree-degree correlations gives the optimal 
robustness of connectivity even for a SF network 
with a power-law degree distribution \cite{Schneider11}\cite{Tanizawa12}.
The name of onion-like comes from 
that it is visualized by the correlations 
when similar degree nodes are set on a concentric circle 
arranged in decreasing order of degrees from core to peripheral. 
Onion-like structure can be generated by 
whole rewiring \cite{Schneider11}\cite{Holme11} 
in enhancing the correlations under a given degree distribution.
On the other hand, 
since dismantling and decycling problems are asymptotically 
equivalent 
at infinite graphs in a large class of random networks 
with light-tailed degree distribution \cite{Braunstein16}, 
trees remain without loops at the critical state before 
the complete fragmentation by node removals.
Dismantling (or decycling) problem known as NP-hard \cite{Karp72} 
is to find the minimum set of nodes in which removal leaves 
a graph broken into connected components 
whose maximum size is at most a constant (or a graph without loops). 
It is suggested from the equivalence that 
the robustness becomes stronger as many loops exist as possible.
In fact, to be the optimal onion-like networks 
at the same level to the rewired ones \cite{Holme11}, 
enhancing loops by copying \cite{Hayashi14} or intermediation 
\cite{Hayashi18a}\cite{Hayashi18b} 
is effective for improving the robustness 
in incrementally growing methods based on a local distributed process
as self-organization. 
Similar effect is also obtained in preserving or non-preserving 
the degrees at nodes 
after the other rewiring based on enhancing loops instead of correlations
\cite{Chujyo20}. 
Thus, we remark that 
loops make bypasses and may be more important 
than the degree-degree correlations 
in order to improve the connectivity in a network reconstruction 
after large disasters or attacks.
It is predicted as the top priority to maximize the decycling set 
(or called Feedback Vertex Set (FVS) in computer science \cite{Karp72}) 
so as not to be tree 
without loops as possible even by the worst case of node removals.
In other words, 
enhancing loops correspond to optimizing the tolerance of connectivity 
in graphs (but not in the contents of general computing or problem 
solving). 
Off course, increasing the path lengths between nodes 
and wasteful resource should be avoided in the reconstruction by 
healing.
However, even 
identifying the necessary nodes to form loops is intractable 
due to combinatorial NP-hardness \cite{Karp72}, 
we effectively apply an approximate calculation by Belief Propagation (BP) 
based on a statistical physics approach in our self-healing through 
rewirings (or additional investment instead) as mentioned later.
We describe the healing methods as sequential processes for computer 
simulation in envisioning the further development of distributed control 
algorithms.

\section{Methods}

\subsection{Outline of Healing Process}
Almost simultaneously attacked nodes are not recoverable immediately, 
therefore are removed from the network function for a while or long time.
In such case of emergency for healing, 
unconnected two nodes are chosen and rewired as the 
reconstruction assistance or reuse of links 
emanated from removed $q N$ nodes, 
when the fraction of attacks is $q$ and 
$N$ denotes the total number of nodes (as the network size). 
Some of disconnected links may be reusable at the neighbor's sides
according to the damage level.
Although we call the reuse rewiring,
removal of nodes is a different problem setting to that 
in the so-called usual rewiring methods 
\cite{Schneider11}\cite{Holme11}\cite{Chujyo20}. 
The outline of healing process is as follows.

\begin{description}
  \item[Step0:] Detection and initiation\\
    After detecting a removal as malfunction at a nearest neighbor of 
    the attacked node, the healing process is initiated autonomously. 
  \item[Step1:] Selection of two nodes\\
    Since the neighbor loses links at least temporary before rewiring, 
    the damaged one becomes an attached candidate for healing.
    Thus, unconnected two nodes are chosen from neighbors of removed 
    nodes by attacks. 
    The selections are different in our proposed and the conventional
    healing methods. Moreover, neighbors are extended in our proposed methods. 
  \item[Step2:] Rewiring for healing\\
    The chosen two nodes are connected as rewiring for healing. 
    The above process is repeated for 
    $M_{h} \stackrel{\rm def}{=} r_{h} \times \tilde{\sum}_{i \in D_{q}} k_{i}$ 
    links. 
\end{description}
Here, 
$\tilde{\sum}_{i \in D_{q}} k_{i}$ means the number of disconnected
links by attacks without multiple counts. 
$D_{q}$ denotes the set of removed nodes, $| D_{q} | = q N$. 
$M_{h}$ includes the number of reused and additionally invested links. 
When reusable links are insufficient, 
we assume to add links as investment 
until to the considered $M_{h}$ for a parameter $0 < r_{h} \leq 1$ 
in computer simulation.

In the healing process, 
rewirings are performed by changing directions and ranges 
of flight routes or wireless beams, 
though we do not discuss the detail realization 
that depends on the current or future technologies and target systems.
We focus on the connectivity at the most fundamental level in many 
network systems for not only communication but also transportation, 
power-supply, and other infrastructures, while our approach may be useful 
for path control or failure detection e.g. by software-defined 
network based reconfiguration on communication systems with switches 
in managing reliability, latency, or security at some service levels
\cite{Song17}\cite{Hu13}\cite{Thyagaturu16}. 
In addition, we consider that ports work independently from links, 
as similar to a relation of airport runaway or plug socket 
and flight by airplane or cable line.
It is reasonable assumption that the amount of degree 
$k_{j}$ ports are reusable at the undamaged neighbor node 
$j \in \partial i$ of a removed node $i$ by attacks, 
where $\partial i$ denotes a set of the nearest connecting neighbors of $i$.
Thus, there exist active (reusable) ports of a node 
at least as many as its degree in the original network before 
attacks.

\subsection{Proposed Healing Methods} \label{subsec1}
Basically, in our proposed healing methods, 
there are two phases: ring formation and enhancing loops by 
applying BP in the next subsection. 
Moreover, they (RingRecal, RingLimit1,5,10, RingLimit5Recal) 
are modified to reduce the additional ports from the previous 
results \cite{Hayashi20} by avoiding the concentration 
of links at some nodes.

\begin{figure}[htb]
\centering
  \includegraphics[width=.73\textwidth]{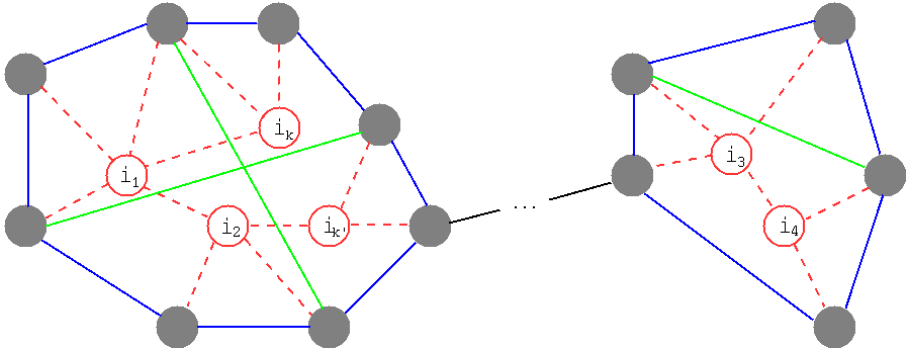}
\caption{Schematic illustration for ring formation and enhancing loops. 
Red nodes and their links are removed by attacks. 
Gray filled nodes are the neighbors. 
Blue lines make rings, and green lines are rewirings 
for enhancing loops on rings.}
\label{fig_ring_loops}
\end{figure}

\begin{description}
  \item[RingBP] 
    Previous our combination method of ring formation 
    and enhancing loops \cite{Hayashi20}.
    After making rings on the extended neighbors of removed nodes 
    as shown in Figure \ref{fig_ring_loops}, 
    enhancing loops on the rings is performed by applying 
    the BP algorithm \cite{Zhou13} (see subsection \ref{subsec2}). 
    However, in the BP, 
    a set of $\{ p_{i}^{0} \}$ as probability of node $i$ 
    to be necessary for loops 
    is calculated only once just after attacks. 
    Note that a ring is the simplest loop by using the least 
    number of links. 
  \item[RingRecal] 
    Modified our method with recalculations of BP. 
    After making rings, a set of $\{ p_{i}^{0} \}$ is recalculated 
    one-by-one through each rewiring in the remaining links within $M_{h}$ 
    for enhancing loops. 
  \item[RingLimit1,5,10] 
    Modified our method with limited rewirings. 
    After making rings, in enhancing loops, 
    the number of rewiring links is limited at node $i$ to 
    its degree $k_{i}$ $+1$, $+5$, or $+10$.
  \item[RingLimit5Recal] 
    Modified our method by a combination of 
    RingRecal and RingLimit5. 
    After making rings, a set of $\{ p_{i}^{0} \}$ is recalculated 
    one-by-one through each rewiring in the remaining links within $M_{h}$
    for enhancing loops. 
    Moreover, 
    the number of rewiring links is limited at node $i$ to $k_{i}+5$.
\end{description}

First, in ring formation (see Figure \ref{fig_ring_loops}), 
the order of process is basically according to the order of 
the removed nodes $i_{1}, i_{2}, \ldots, i_{qN}$.
Thus, rings are made for the neighbors 
$\partial i_{1}, \partial i_{2}, \ldots, \partial i_{qN}$ 
in this order. 
However, if there is $i_{k'} \in \partial i_{k}$, $k' > k$, 
it is extended as the union 
$\partial i_{k} \leftarrow \partial i_{k} \cup \partial i_{k'}$. 
In addition, 
if there is $i_{k''} \in \partial i_{k'}$, $k'' > k'$, 
it is also extended as the union
$\partial i_{k} \leftarrow \partial i_{k} 
\cup \partial i_{k'} \cup \partial i_{k''}$. 
Such extensions of neighbors are repeated until that 
a ring encloses the induced subgraph of removed nodes and their links.
To make a ring, 
a node is chosen u.a.r and connect to a subsequent similarly chosen node 
in a set of the extended neighbors. This is repeated without multi-selections 
until return to the first chosen node from the last chosen node.

Next, 
in enhancing loops on each ring for remained rewirings in $M_{h}$, 
a node $j$ with the minimum $p_{j}^{0}$ is chosen in all of the 
neighbors of removed nodes, 
and connected to other node $j'$ with the second minimum $p_{j'}^{0}$ 
on the ring to which $j$ belongs.
For each rewiring, 
a set of $\{ p_{i}^{0} \}$ is recalculated one-by-one in RingRecal 
and RingLimit5Recal methods. 
In addition, 
the number of rewiring links is limited at node $i$ to $k_{i}+5$
(or $+1$, $+10$) according to its degree $k_{i}$ in 
RingLimit5Recal and RingLimit5 (or RingLimit1, RingLimit10) methods. 
If the condition is unsatisfied, 
other node with the second, third, forth, and subsequent minimum
is chosen as a candidate for healing.
Although 
a node $j''$ with small $p_{j''}^{0}$ tends to not contribute to 
making loops because of not included in FVS, 
it is expected that the number of loops is increased by 
connecting such nodes. This is the reason for the above selection.

\subsection{Applying Belief Propagation Algorithm} \label{subsec2}
To calculate the probability $p_{i}^{0}$ of belonging to FVS, 
the following BP algorithm \cite{Zhou13} are applied. 
It is based on a cavity method in statistical physics.
We review the outline derived for approximately estimating FVS 
known as NP-hard problem \cite{Karp72}.
In the cavity graph, it is assumed that nodes $j \in \partial i$ 
are mutually independent of each other when node $i$ is removed 
(the exception is denoted by $\backslash i$).
Then the joint probability is 
${\cal P}_{\backslash i}({A_{j}: j \in \partial i}) 
  \approx \Pi_{j \in \partial i} p^{A_{j}}_{j \rightarrow i}$ 
by the product of independent marginal probability 
$p^{A_{j}}_{j \rightarrow i}$ 
for the state $A_{j}$ as the node index of $j$'s root or 
empty $0$: it belongs to FVS.
The corresponding probabilities are represented by 
\begin{equation}
  p^{0}_{i} \stackrel{\rm def}{=} \frac{1}{z_{i}(t)}, \label{eq_BP1}
\end{equation}
\begin{equation}
  p^{0}_{i \rightarrow j} = \frac{1}{z_{i \rightarrow j}(t)}, 
\end{equation}
\begin{equation}
  p^{i}_{i \rightarrow j} = 
  \frac{e^{x} \Pi_{k \in \partial i(t)\backslash j} 
    \left[ p^{0}_{k \rightarrow i} + p^{k}_{k \rightarrow i} \right]
  }{z_{i \rightarrow j}(t)}, 
\end{equation}
where $\partial i(t)$ denotes node $i$'s set of connecting neighbor nodes 
at time $t$, 
and $x > 0$ is a parameter of inverse temperature. 
The normalization constants are 
\begin{equation}
  z_{i}(t) \stackrel{\rm def}{=} 
  1 + e^{x} \left[ 1 + \sum_{k \in \partial i(t)} 
      \frac{1 - p^{0}_{k \rightarrow i}}{
        p^{0}_{k \rightarrow i} + p^{k}_{k \rightarrow i}} \right]
  \Pi_{j \in\partial i(t)} 
  \left[ p^{0}_{j \rightarrow i} + p^{j}_{j \rightarrow i} \right],
  \label{eq_BP4}
\end{equation}
\begin{eqnarray}
  z_{i \rightarrow j}(t) & \stackrel{\rm def}{=} &
  1 + e^{x} \Pi_{k \in \partial i(t)\backslash j} 
  \left[ p^{0}_{k \rightarrow i} + p^{k}_{k \rightarrow i} \right] 
  \times \left[ 1 + \sum_{l \in \partial i(t)\backslash j} 
    \frac{1 - p^{0}_{l \rightarrow i}}{
      p^{0}_{l \rightarrow i} + p^{l}_{l \rightarrow i}} \right], 
  \label{eq_BP5}
\end{eqnarray}
to be satisfied for any node $i$ and link $i \rightarrow j$ as 
\[
  p^{0}_{i} + p^{i}_{i} + \sum_{k \in \partial i} p^{k}_{i} = 1, \;\;\; 
  p^{0}_{i \rightarrow j} + p^{i}_{i \rightarrow j} 
  + \sum_{k \in \partial i} p^{k}_{i \rightarrow j} = 1.
\]
We repeat these calculations of message-passing 
until to be self-consistent in principle 
but practically to reach appropriate rounds 
from initial setting of $(0,1)$ random values.
The unit time from $t$ to $t+1$ for calculating a set 
$\{ p^{0}_{i} \}$ consists of a number of rounds by 
updating equations (\ref{eq_BP1})-(\ref{eq_BP5}) 
in order of random permutation of the total $N$ nodes.
Since the sums or products in equations (\ref{eq_BP1})-(\ref{eq_BP5}) 
are restricted in the nearest neighbor, 
they are local processes. 
The distributed calculations can be also considered.
As included in FVS, 
a node $k$ with the maximum $p_{k}^{0}$ is chosen.
After removing the chosen node, $\{ p_{i}^{0} \}$ is recalculated 
at next time. Such process is repeated 
until to be acyclic for finding the FVS.
However, in our healing method, $\{ p_{i}^{0} \}$ is used for 
selecting attached two nodes on a ring by rewiring.

\subsection{Conventional Healing Methods} \label{subsec3}
We briefly explain the following typical 
healing methods in network science (inspired from fractal 
statistical physics) and computer science.
\begin{description}
  \item[RBR] 
    Conventional Random Bypass Rewiring (RBR) method \cite{Park16}
    (corresponded to $r_{h} = 0.5$).
  \item[GBR] 
    Greedy Bypass Rewiring (GBR) method improved from RBR heuristically \cite{Park16}.
  \item[SLR] 
    Conventional Simple Local Repair (SLR) method \cite{Gallos15} with 
    priority of rewirings to more damaged nodes.
\end{description}

In network science, a self-healing method 
by adding new random links on interdependent two-layered 
networks of square lattices has been proposed, 
and the effect against node attacks 
is numerically studied \cite{Kertesz14}.
In particular, for adding links by the healing process, 
the candidates of linked nodes are incrementally extended from 
only the direct (nearest connecting) neighbors of the removed node 
by attacks until no more separation of components occurs. 
In other words, 
the whole connectivity is maintained except the isolating 
removed parts. 
Such extension of the candidates of linked nodes
is a key idea in our proposed self-healing method. 

Furthermore, 
the following self-healing methods, 
whose effects are investigated for some data of real networks, 
are worthy to note.
One is a distributed SLR \cite{Gallos15} 
with the repair by a link between 
the most damaged node and a randomly chosen node from the 
unremoved node set in its next-nearest neighbors before attacks.
The priority of damaged nodes is according to 
the smaller fraction $k_{dam} / k_{orig}$ of its remained degree $k_{dam}$ 
and the original degree $k_{orig}$ before the attacks.
The selections are repeated 
until reaching a given rate $f_{s}$ controlled by 
the fraction of nodes whose $k_{dam} / k_{orig}$ falls bellow a threshold. 
Another is RBR \cite{Park16} 
on more limited resource of links and ports. 
To establish links between pair nodes, 
a node is randomly chosen only one time in the neighbors of 
each removed node. 
When $k_{i}$ denotes the degree of removed node $i$, 
only $\lfloor k_{i}/2 \rfloor$ links are reused.
Note that 
reserved additional ports are not necessary: 
they do not exceed the original one before attacks. 
Moreover, GBR \cite{Park16} is proposed in order to improve the 
robustness, the selection of pair nodes is based on the number 
of the links not yet rewired and the size of the neighboring components.

In computer science, 
ForgivingTree algorithm has been proposed \cite{Hayes08}.
Under the repeated attacks, 
the following self-healing is processed one-by-one after each 
node removal, except when the removed node is a leaf 
(whose degree is one).
It is based on both distributed process of sending messages 
and data structure, furthermore developed to an efficient 
algorithm called as compact routing \cite{Castaneda18}. 
In each rewiring process, 
a removed node and its links are replaced by a binary tree. 
Note that each vertex of the binary tree was the neighbors 
of the removed node, whose links to the neighbors are reused 
as the edges of the binary tree. 
Thus, additional links for healing is unnecessary, 
but not controllable.
It is remarkable for computation 
(e.g., in routing or information spreading) that 
the multiplicative factor of diameter of the graph after healing 
is never more than 
$O( \log k_{max})$, where $k_{max}$ is the maximum degree in 
the original network because of the replacing by binary trees.
However, 
the robustness of connectivity is not taken into account 
in the limited rewiring based on binary trees, 
since a tree structure is easily fragmented into subtrees 
by any attack to the articulation node.
Thus, this healing method is excluded from compared ones.

\section{Results}
We evaluate the effect of healing by four measures: 
the ratio $S(q)/N_{q}$ \cite{Gallos15} for the connectivity, 
the robustness index 
$R(q) \stackrel{\rm def}{=} \sum_{q'} S_{q}(q')/N_{q}$, 
the efficiency of paths 
$E(q) \stackrel{\rm def}{=} \frac{1}{N_{q}(N_{q}-1)} 
\sum_{i \neq j} \frac{1}{L_{ij}}$, 
and the average degree $k_{avg}(q)$ in 
$N_{q} \stackrel{\rm def}{=} (1-q)N$ nodes, 
where $S(q)$ and $S_{q}(q')$ denote 
the sizes of GC (giant component or largest connected cluster) 
after removing $q N$ nodes by attacks from the original network 
and removing $q' N_{q}$ nodes by further attacks from the surviving 
$N_{q}$ nodes, respectively.
Here, a removed node is chosen 
with recalculation of the highest degree node as the target.
Remember that $q = 1/N, 2/N, \ldots, (N-1)/N$ 
(or $q' = 1/N_{q}, 2/N_{q}, \ldots, (N_{q}-1)/N_{q}$) 
is a fraction of attacks.
While $S(q)$ or $S_{q}(q')$ represents the size of GC after attacks to 
$q N$ or $q' N_{q}$ nodes, 
$R(q)$ is a measure of tolerance of connectivity against further attacks. 
$L_{ij}$ denote 
the length of the shortest path counted by hops between $i$-$j$ nodes 
in the surviving $N_{q}$ nodes. 
The ranges are $0 < S(q)/N_{q} \leq 1$, $0 < R(q) \leq 0.5$, and 
$0 < E(q) \leq 1$. 
We investigate the four measures before or after healing 
for OpenFlights between airports, Internet AS Oregon, 
and US PowerGrid 
as examples of typical infrastructure of SF networks \cite{net_data} 
after extracting from each 
of them to a connected and undirected graph without multiple links 
(see Table \ref{table_data}). 
We compare the results shown by color lines with marks in figures 
for the conventional RBR, GBR, SLR, 
and our proposed RingBP, RingRecal, RingLimit1,5,10, 
RingLimit5Recal methods. 

\begin{figure}[htb]
\centering
  \includegraphics[width=\textwidth]{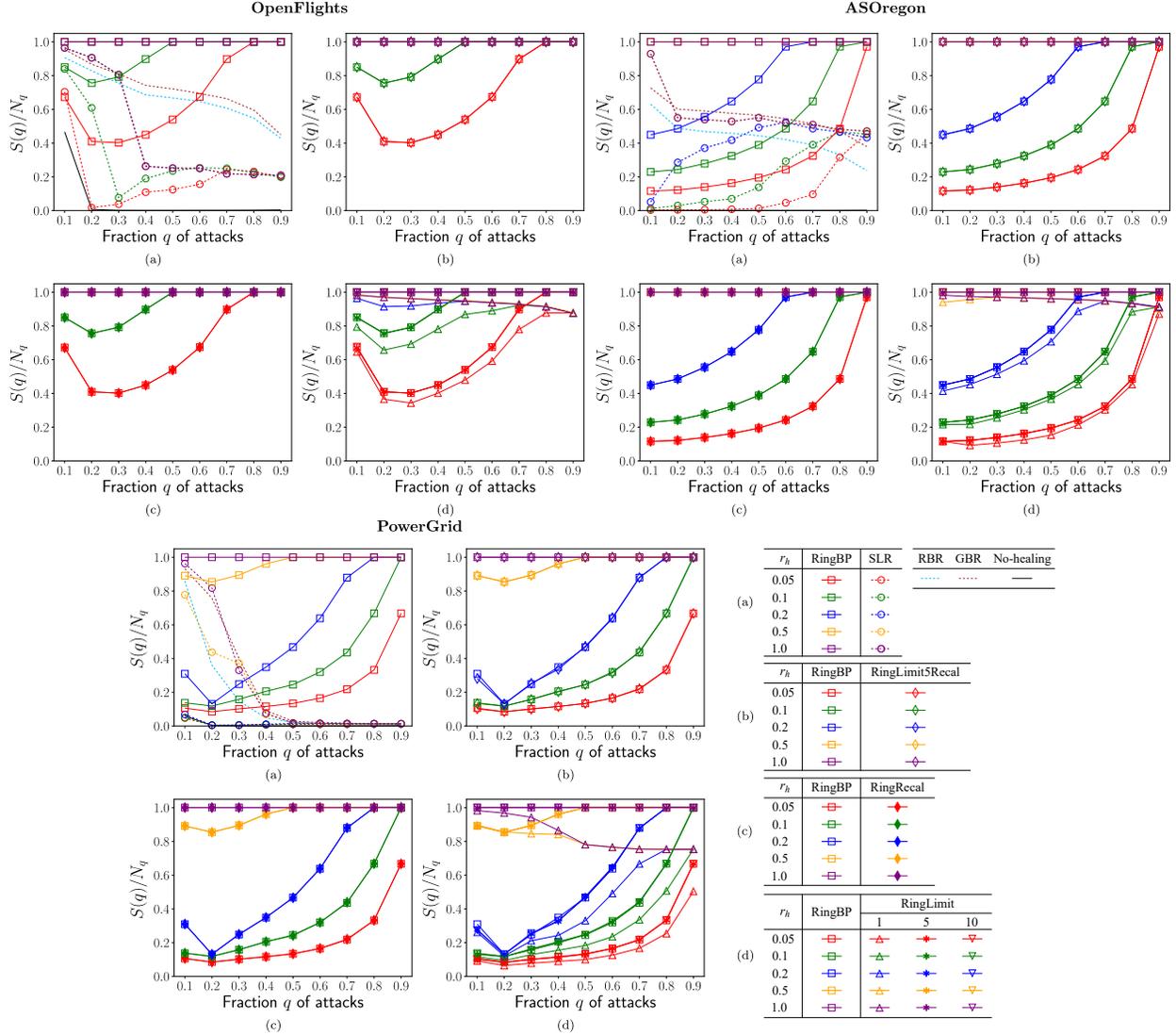}
\caption{Ratio $S(q)/N_{q}$ of connectivity vs fraction $q$ of attacks 
for the rate $r_{h}$ in rewirings.}
\label{fig_connect}
\end{figure}

\begin{table}[htb]
\caption{Basic properties for the original networks. 
$N$ and $M$ denote the numbers of nodes and links.
$k_{avg} = 2 M / N$, 
$k_{min}$, and $k_{max}$ are the average, minimum, and maximum degrees.
$L_{avg}, D, R$ and $E$ denote the average path length, diameter, robustness index, 
and efficiency of paths, respectively.}
\label{table_data}
\begin{footnotesize}
\centering
\begin{tabular}{c|ccccccccc} \hline
{\bf Network} & {\bf $N$} & {\bf $M$} & {\bf $k_{avg}$} & {\bf $k_{min}$} 
& {\bf $k_{max}$} & {\bf $L_{avg}$} & {\bf $D$} & {\bf $R$} & {\bf $E$} \\ \hline
OpenFlight & 2905 & 15645 & 10.77 & 1 & 242 & 4.097  & 14 & 0.080912 & 0.266934\\
AS Oregon  & 6474 & 12572 & 3.883 & 1 & 1458 & 3.705 & 9  & 0.012500 & 0.290399\\
PowerGrid  & 4941 & 6594  & 2.669 & 1 & 19  & 18.989 & 46 & 0.052428 & 0.062878\\
\hline
\end{tabular}
\end{footnotesize}
\end{table}

\begin{figure}[htb]
\centering
  \includegraphics[width=\textwidth]{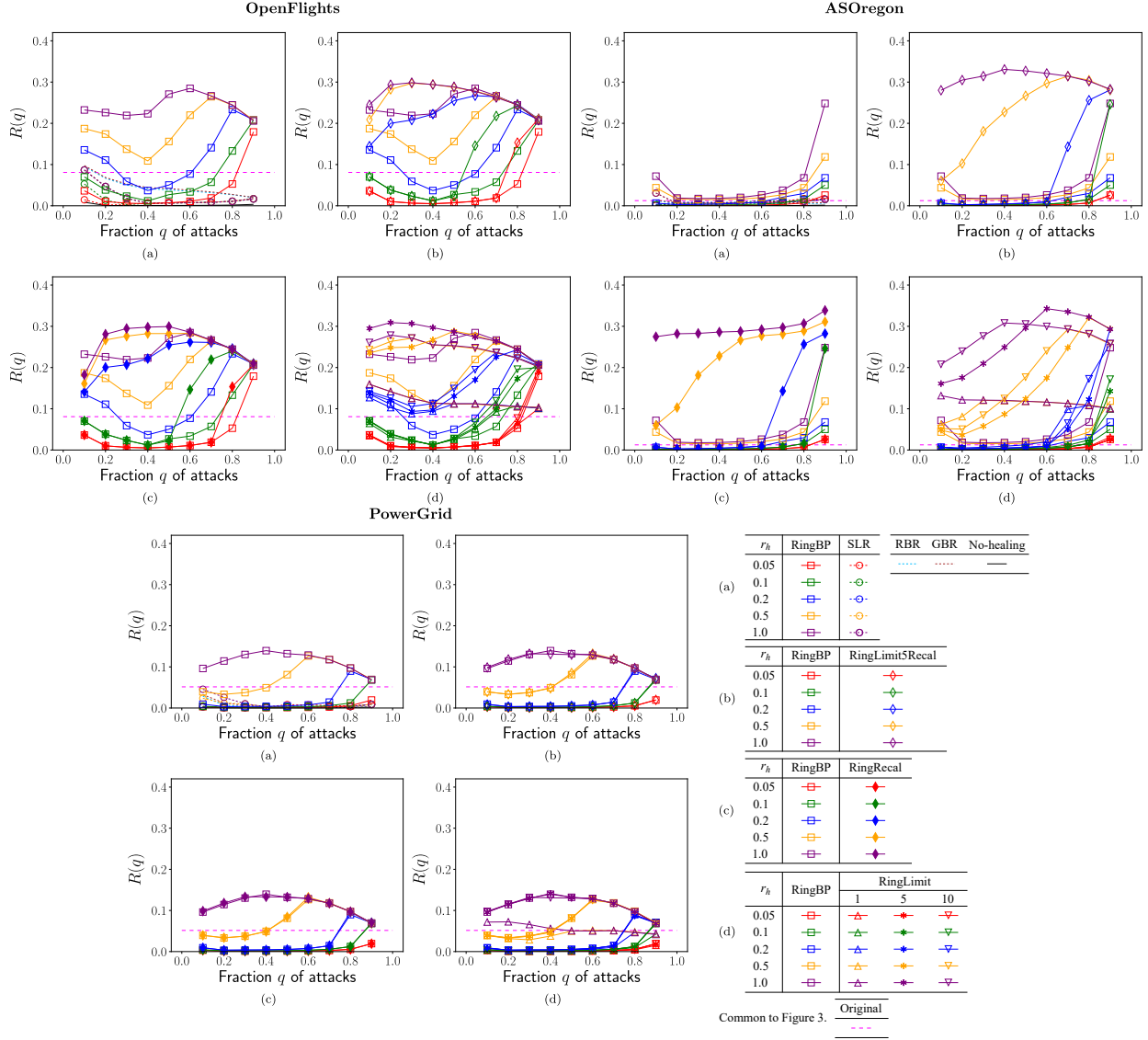}
\caption{Robustness index as the tolerance of connectivity against further 
attacks to the surviving $N_{q}$ nodes vs fraction $q$ of attacks 
for the rate $r_{h}$ in rewirings.}
\label{fig_robust}
\end{figure}

\begin{figure}[htb]
\centering
  \includegraphics[width=\textwidth]{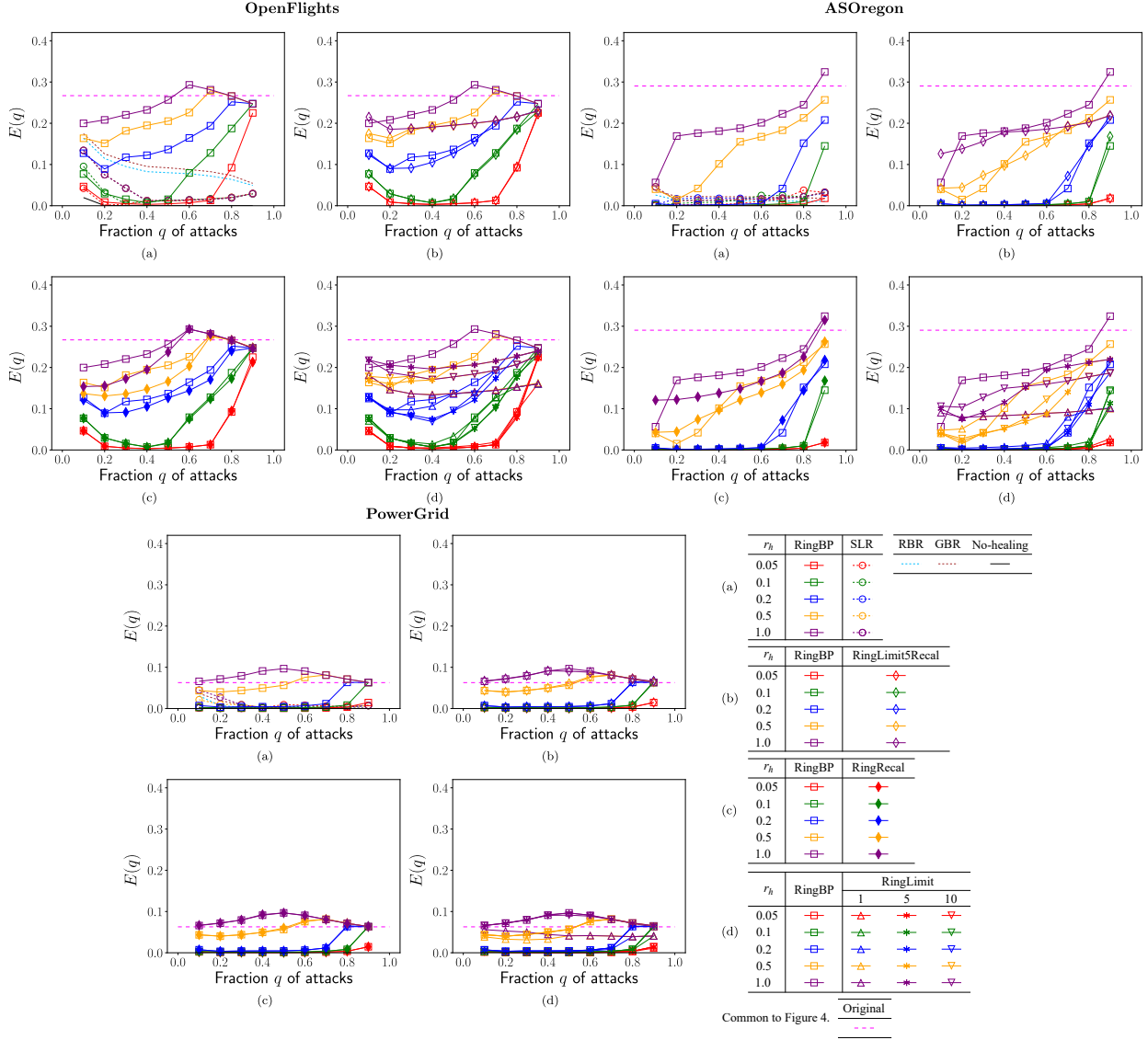}
\caption{Efficiency of paths in the surviving $N_{q}$ nodes 
vs fraction $q$ of attacks  
for the rate $r_{h}$ in rewirings.}
\label{fig_eff}
\end{figure}

\begin{figure}[htb]
\centering
  \includegraphics[width=\textwidth]{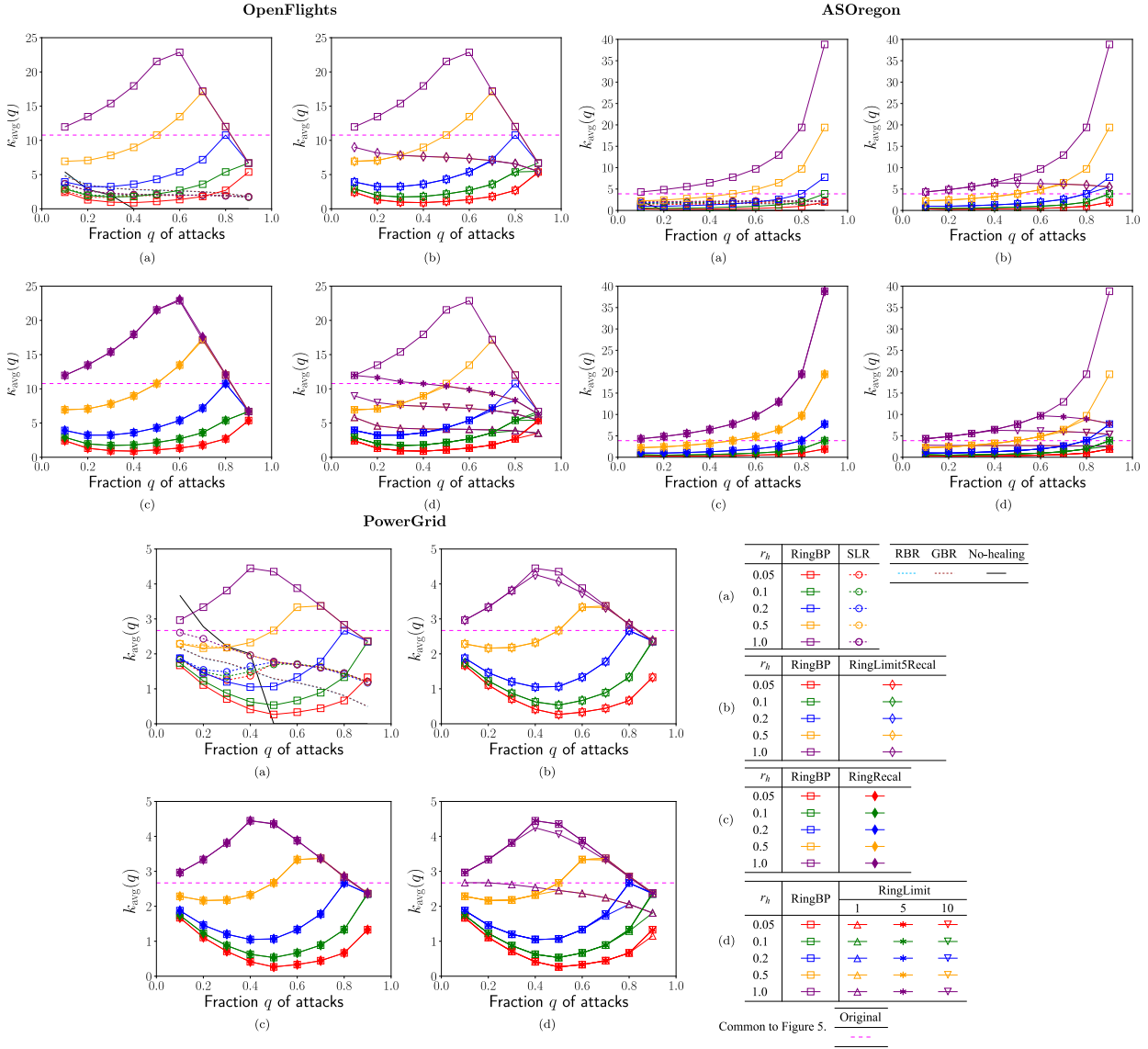}
\caption{Average degree $k_{avg}(q)$ in the surviving $N_{q}$ nodes 
vs fraction $q$ of attacks 
for the rate $r_{h}$ in rewirings.} 
\label{fig_ave_deg}
\end{figure}

In each Figure \ref{fig_connect},\ref{fig_robust},\ref{fig_eff},\ref{fig_ave_deg},
no-healing, conventional, and previous our methods are compared in (a), 
previous our and RingRecal or RingLimit methods are compared in (c)(d), 
previous our and the best combination 
RingLimit5Recal methods are compared in (b). 
Red, green, blue, orange, and purple lines denote the rate 
$r_{h} = 0.05, 0.1, 0.2, 0.5$ and $1.0$, respectively, 
for the number $M_{h}$ of rewirings. 
The results for the original and no-healing networks are shown by 
dashed magenta and solid black lines. 
The following results are averaged over $100$ samples with 
random process for tie-breaking in a node selection or 
ordering of nodes on a ring.

Figure \ref{fig_connect} shows the ratio $S(q)/N_{q}$ of connectivity in 
the surviving $N_{q}$ nodes.
Remember that $S(q)$ is the size of GC after healing (or no-healing) 
against attacks to $q N$ nodes. 
Higher ratio means larger connectivity as maintaining the 
network function for communication or transportation, 
$S(q)/N_{q} < 1$ indicates the incomplete ring formation stopped in $M_{h}$. 
As shown in Figure \ref{fig_connect}(a), 
the ratio rapidly decreases in the conventional SLR method 
marked by open circles for OpenFlights and PowerGrid, 
while it is moderately higher around $S(q)/N_{q} \approx 0.5$ on purple 
and blue lines or increasing in green and red lines marked by open circles 
for AS Oregon.
Moreover, in Figure \ref{fig_connect}(a), 
the following results are common for OpenFlights, AS Oregon, and PowerGrid. 
The ratio also decreases in the conventional RBR and GBR methods 
denoted by dashed light-blue and brown lines, respectively. 
In the corresponding RingBP method, 
the ratio is the highest as the horizontal orange (but overlapped purple) 
line at $S(q)/N_{q} \approx 1.0$ marked by open squares.
The bottom dashed black lines around $S(q)/N_{q} = 0$ are 
the results without the network function for no-healing. 
Thus, previous our RingBP method marked by open squares 
has higher ratio than the conventional methods 
in comparison with same color lines. 
Figure \ref{fig_connect}(b)(c) shows that the ratio in RingBP method 
marked by open squares almost coincide with ones in RingLimit5Recal method 
marked by open diamonds and RingRecal method marked by filled diamonds.
Similarly, Figure \ref{fig_connect}(d) shows that 
the ratio in RingBP method marked by open squares almost coincide 
with ones in RingLimit5,10 methods marked by lower triangles and 
asterisks.
However it is slightly lower in RingLimit1 method marked by open upper 
triangles. 
Therefore, RingLimit5Recal, RingRecal, and RingLimit5,10 methods are 
the best at the same level to RingBP in maintaining the connectivity. 
The constraint to the number of additional ports is slightly too 
strong as only one in RingLimit1 method.

Figure \ref{fig_robust} shows the robustness index $R(q)$ as the tolerance 
of connectivity against further attacks to the surviving $N_{q}$ nodes 
after healing. 
Note that a major part of $N_{q}$ nodes belong to the GC 
but other parts belong to isolated clusters. 
In Figure \ref{fig_robust}(a) for OpenFlights, AS Oregon, and PowerGrid, 
the values of $R(q)$ rapidly decrease to very low level $\leq 0.1$ with 
vulnerability in the conventional SLR method marked by open circles 
and in RBR and GBR methods denoted by light-blue and brown dashed lines, 
while there exist higher values of $R(q)$ 
(on purple and orange lines for $r_{h} \geq 0.5$) 
in RingBP method marked by open squares than 
the horizontal dashed magenta lines in the original network.
The results for no-healing are at the bottom as $R(q) \approx 0$ 
because of $S_{q} \approx 0$ from Figure\ref{fig_connect}(a). 
Moreover, as shown in Figure \ref{fig_robust}(b)(c), 
RingLimit5Racal method marked by open diamonds and 
RingRecal method marked by filled diamonds have higher values of $R(q)$ 
than RingBP marked by open squares in comparison with same color lines 
for OpenFlights and AS Oregon, 
while these methods have almost same values of $R(q)$ to ones in 
RingBP for PowerGrid. 
Similarly, 
as shown in Figure \ref{fig_robust}(d) for OpenFlights and AS Oregon, 
RingLimit1,5,10 methods marked by open lower, upper triangles and asterisks 
have higher values of $R(q)$ than RingBp method marked by open squares 
in comparison with same color lines. 
However, the difference becomes smaller in green and red lines 
for $r_{h} \leq 0.1$. 
In Figure \ref{fig_robust}(d) for PowerGrid, 
similar values of $R(q)$ are obtained on each color lines 
regardless of marks for different methods.
Partially, for OpenFlights and AS Oregon, 
purple, orange and blue line ($r_{h} \geq 0.2$) in RingLimit5Recal 
marked by open diamonds are slightly higher than ones in Ringlimit1,5,10 
marked by open upper, lower triangles and asterisks 
as shown in Figure \ref{fig_robust}(b)(d). 
Thus, the reconstructed networks by our proposed healing methods 
can become more stronger with higher values of $R(q)$ 
than the original network against further attacks. 
In particular, the improvement is remarkable 
from $R(q) < 0.1$ to $R(q) > 0.3$ for OpenFlights and AS Oregon.

Figure \ref{fig_eff} shows the efficiency $E(q)$ 
of shortest paths between two nodes in the surviving $N_{q}$ nodes. 
Note that 
$E(q) = 0.1, 0.2, 0.25$ is corresponded to $10, 5, 4$ hops of the average 
path length $L_{avg}(q)$ from $L_{avg}(q) \approx 1/E(q)$ 
in the arithmetic and the harmonic means of path lengths.
The following results are common for OpenFlights, AS Oregon, and PowerGrid. 
As similar to Figures \ref{fig_connect}(a) and \ref{fig_robust}(a), 
Figure \ref{fig_eff}(a) shows that 
the values of $E(q)$ rapidly decrease in the conventional SLR method 
marked by open circles, RBR and GBR methods denoted by light-blue and 
brown dashed lines, while the values are higher in RingBP method 
marked by squares in comparison with same color lines.
In Figure \ref{fig_eff}(b)(c), 
RingLimit5Recal method marked by open diamonds 
and RingRecal method marked by filled diamonds have similar or slightly 
lower values of $E(q)$ than ones in RingBP method marked by squares 
in comparison with same color lines. 
In Figure \ref{fig_eff}(d) for OpenFlights and AS Oregon, 
the values are slightly lower in RingLimit1,3,5 methods marked by open upper, 
lower triangles and asterisks than ones in RingBP method marked by squares, 
while for PowerGrid 
the values are similar regardless of these methods 
in comparison with same color lines.

\begin{figure}
\centering
  \includegraphics[width=.9\textwidth]{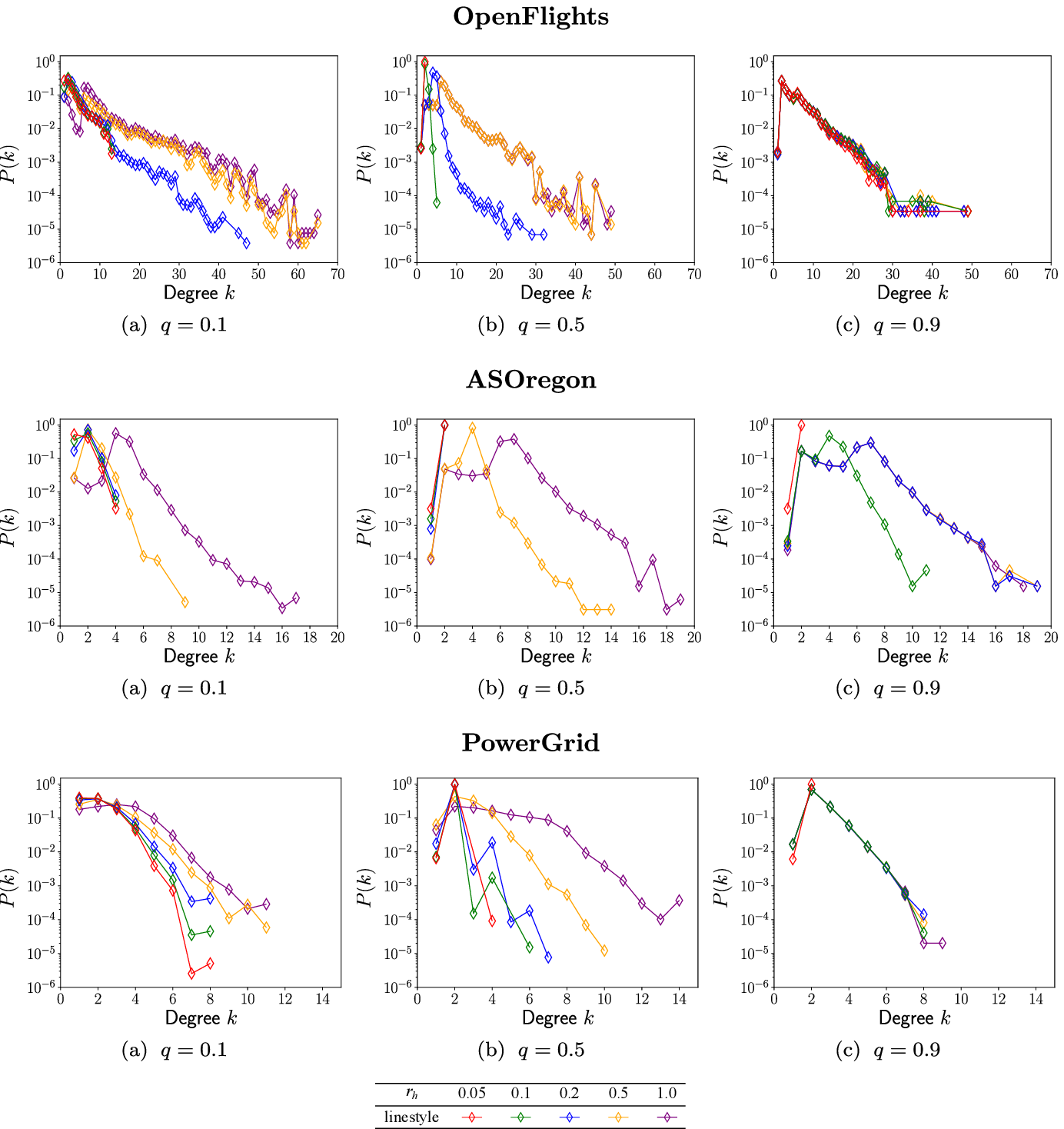}
\caption{Degree distribution in surviving $N_{q}$ nodes 
after healing by RingLimit5Recal method for the fraction of attacks.}
\label{fig_deg_dist}
\end{figure}

Figure \ref{fig_ave_deg} shows 
the average degree $k_{avg}(q)$ in the surviving $N_{q}$ nodes. 
This value indicates how much links are effectively used for hearing.
In other words, 
a small value of $k_{avg}(q)$ means that rewirings are restricted 
and not fully used until the possible number 
$M_{h}$ of links especially in the conventional methods, 
by the constraints on linking between not the extended but 
the nearest neighbors of attacked nodes or the limitation
(see the subsection \ref{subsec3}). 
The following results are common for OpenFlights, AS Oregon, and 
PowerGrid.
As shown in Figure \ref{fig_ave_deg}(a), 
it is remarkable that the values are small $k_{avg}(q) < 10$ 
in the conventional SLR method marked by open circles, RBR and GBR methods 
denoted by dashed light-blue and brown lines, 
while the values are higher 
in RingBP method marked by open squares in comparison with same color lines.
In Figure \ref{fig_ave_deg}(b)(d), 
by saving rewired links due to the limitation of additional ports, 
the values of $k_{avg}(q)$ are not large 
in RingLimit5Recal method marked by open diamonds or 
in RingLimit1,5,10 methods marked by open upper, lower triangles and 
asterisks. 
In Figure \ref{fig_ave_deg}(c), on each color line, 
the values of $k_{avg}(q)$ in RingBP method marked by open squares are 
almost coincident with ones in RingRecal method marked by filled diamonds. 
However, 
in RingLimit5Recal and RingLimit1,5,10 methods with saving rewired links, 
both $R(q)$ and $E(q)$ are high values 
as shown in Figures \ref{fig_robust}(b)(d) and \ref{fig_eff}(b)(d). 
Therefore, these methods are more effective for healing to improve the robustness 
and efficiency to similar levels by using less resource. 
Figure \ref{fig_deg_dist} shows that 
the reconstructed degree distribution $P(k)$ in RingLmit5Recal method 
becomes exponential approximately in a semi-logarithmic plot 
from a power-law in the original network.
The maximum degrees are bounded as $65$, $19$, and $14$ 
for OpenFlights, AS Oregon, and PowerGrid, respectively. 
They tend to be smaller as $q$ increases.

\begin{table}
\caption{Maximum number of additional ports 
(the average number in parenthesis) for the fraction $q$ of attacks and 
the rate $r_{h}$ in rewirings.}
\label{table_add_ports}
\begin{tiny}
\centering
OpenFlights

\begin{tabular}{c|ccccccccc} \hline
\backslashbox{$r_{h}$}{$q$} 
& 0.1  & 0.2  & 0.3  & 0.4  & 0.5  & 0.6  & 0.7  & 0.8  & 0.9 \\ \hline \hline
0.05 & 1 & 1 & 1 & 1 & 1 & 1 & 1 & 2 & 14.7 \\
& (1) & (1) & (1) & (1) & (1) & (1) & (1) & (1.36) & (3.32) \\ \hline
0.1 & 1 & 1 & 1 & 1 & 2 & 2 & 4.03 & 20.68 & 21.59 \\
& (1) & (1) & (1) & (1) & (1.10) & (1.37) & (1.91) & (3.55) & (4.46) \\ \hline
0.2 & 26.54 & 3.67 & 3.44 & 3.81 & 10.22 & 29.44 & 44.08 & 49.23 & 21.56 \\
& (1.66) & (1.59) & (1.68) & (1.91) & (2.51) & (3.64) & (5.47) & (8.46) & (4.46) \\ \hline
0.3 & 55.44 & 23.93 & 26.55 & 42.07 & 50.64 & 65.58 & 69.37 & 63.04 & 21.61 \\
& (3.57) & (2.83) & (3.02) & (3.72) & (4.86) & (6.56) & (8.92) & (9.74) & (4.46) \\ \hline
0.4 & 77.65 & 56.61 & 62.22 & 69.58 & 79.96 & 88.11 & 81.62 & 63.09 & 21.6 \\
& (5.57) & (4.44) & (4.81) & (5.74) & (7.20) & (9.28) & (12.17) & (9.73) & (4.46) \\ \hline
0.5 & 96.35 & 81.88 & 87.52 & 98.26 & 106.62 & 105.12 & 107.14 & 62.99 & 21.62 \\
& (7.56) & (6.07) & (6.61) & (7.77) & (9.59) & (11.84) & (14.84) & (9.74) & (4.46) \\ \hline
1.0 & 179.04 & 175.97 & 173.53 & 177.32 & 160.94 & 170.93 & 115.33 & 63.02 & 21.57 \\
& (16.78) & (13.95) & (14.85) & (16.91) & (19.68) & (20.66) & (15.10) & (9.74) & (4.46) \\ \hline
\end{tabular}
\vspace{3mm}

\centering
AS Oregon

\begin{tabular}{c|ccccccccc} \hline
\backslashbox{$r_{h}$}{$q$} 
& 0.1  & 0.2  & 0.3  & 0.4  & 0.5  & 0.6  & 0.7  & 0.8  & 0.9 \\ \hline \hline
0.05 & 1 & 1 & 1 & 1 & 1 & 1 & 1 & 1 & 1 \\
& (1) & (1) & (1) & (1) & (1) & (1) & (1) & (1) & (1) \\ \hline
0.1 & 1 & 1 & 1 & 1 & 1 & 1 & 1 & 1 & 6.43 \\
& (1) & (1) & (1) & (1) & (1) & (1) & (1) & (1) & (2.26853) \\ \hline
0.2 & 1 & 1 & 1 & 1 & 1 & 1 & 2.03 & 6.44 & 56.47 \\
& (1) & (1) & (1) & (1) & (1) & (1) & (1.29) & (2.23) & (6.30) \\ \hline
0.3 & 1 & 1 & 1 & 1 & 2 & 3.08 & 6.67 & 49.06 & 85.74 \\
& (1) & (1) & (1) & (1) & (1.21) & (1.42) & (2.22) & (4.21) & (10.38) \\ \hline
0.4 & 1 & 1 & 2 & 2 & 3.66 & 5.48 & 48.89 & 79.8 & 107.53 \\
& (1) & (1) & (1.16) & (1.31) & (1.57) & (2.21) & (3.53) & (6.24) & (14.48) \\ \hline
0.5 & 2.49 & 2 & 2 & 3.77 & 6.22 & 48.81 & 79.89 & 104.12 & 127.04 \\
& (1.11) & (1.26) & (1.37) & (1.68) & (2.22) & (3.20) & (4.88) & (8.25) & (18.54) \\ \hline
1.0 & 40.64 & 73.63 & 93.29 & 111.4 & 128 & 146.73 & 164.37 & 181.31 & 199.37 \\
& (2.67) & (3.20) & (3.92) & (4.87) & (6.22) & (8.21) & (11.56) & (18.30) & (38.81) \\ \hline
\end{tabular}
\vspace{3mm}

\centering 
PowerGrid

\begin{tabular}{c|ccccccccc} \hline
\backslashbox{$r_{h}$}{$q$} 
& 0.1  & 0.2  & 0.3  & 0.4  & 0.5  & 0.6  & 0.7  & 0.8  & 0.9 \\ \hline \hline
0.05 & 1.08 & 1 & 1.02 & 1.01 & 1 & 1 & 1 & 1 & 1 \\
& (1.00) & (1) & (1.00) & (1.00) & (1) & (1) & (1) & (1) & (1) \\ \hline
0.1 & 1.65 & 1.99 & 1.93 & 1.59 & 1.15 & 1.02 & 1.04 & 1.02 & 3.81 \\
& (1.01) & (1.01) & (1.02) & (1.00) & (1.00) & (1.00) & (1.00) & (1.00) & (1.19) \\ \hline
0.2 & 2 & 2.1 & 2.12 & 2.25 & 2.13 & 1.91 & 2.01 & 3.8 & 3.69 \\
& (1.02) & (1.02) & (1.03) & (1.03) & (1.02) & (1.00) & (1.01) & (1.32) & (1.19) \\ \hline
0.3 & 2 & 2.59 & 2.95 & 2.93 & 2.95 & 3.01 & 3.12 & 6.21 & 3.74 \\
& (1.02) & (1.05) & (1.07) & (1.07) & (1.08) & (1.10) & (1.27) & (1.51) & (1.19) \\ \hline
0.4 & 2.42 & 3.06 & 3.25 & 3.14 & 3.25 & 3.11 & 8.99 & 6.27 & 3.74 \\
& (1.04) & (1.09) & (1.10) & (1.13) & (1.15) & (1.26) & (1.90) & (1.51) & (1.19) \\ \hline
0.5 & 2.87 & 3.76 & 3.47 & 3.47 & 3.31 & 5.66 & 9.03 & 6.35 & 3.77 \\
& (1.06) & (1.11) & (1.14) & (1.17) & (1.27) & (1.78) & (1.90) & (1.51) & (1.19) \\ \hline
1.0 & 4.9 & 6.08 & 13.42 & 15.44 & 14.95 & 11.85 & 9.01 & 6.29 & 3.78 \\
& (1.64) & (1.86) & (2.31) & (2.94) & (2.80) & (2.35) & (1.90) & (1.51) & (1.19) \\ \hline
\end{tabular}
\end{tiny}
\end{table}

Moreover, 
Table \ref{table_add_ports} shows the maximum number (or in parentheses, 
the average value over the nodes that perform much more rewirings 
than their degrees of the reusable number of ports) 
of additional ports in RingRecal method. 
Although the values are reduced to less than $k_{max}$ from nearly 
$2 k_{max} \sim 3 k_{max}$ in previous our RingBP method \cite{Hayashi20}, 
they are still large. 
Here, $k_{max}$ is $242$, $1458$, or $19$ for the original networks:  
OpenFlights, AS Oregon, or PowerGrid as shown in Table \ref{table_data}. 
Off course, the maximum number of additional ports is significantly 
restricted to a constant $1, 5, 10$ or $5$ in RingLimit1,5,10 
or RingLimit5Recal method. 
Since additional ports should be stored in advance beside 
reusable number of its degree in the original network, 
fewer preparing is better within lower investment cost of resource.
Thus, RingBP or RingRecal method is not desirable because of requiring 
many additional ports.

\section{Discussion}
We have proposed self-healing methods with modifications from 
the previous one \cite{Hayashi20} 
for reconstructing a resilient network through 
rewirings against attacks or disasters in resource allocation 
control of links and ports.
The healing strategy is based on maintaining the connectivity 
by ring formation on the extended neighbors of attacked nodes
and enhancing loops for improving the robustness of connectivity
in applying the approximate calculations of BP \cite{Zhou13} 
inspired from statistical physics in distributed manner. 
We have taken into account 
the limitation of additions and the recalculations of BP 
as modifications to reduce the preparing of additional ports 
by avoiding the concentration of links at some nodes.

Simulation results show that 
our proposed combination methods of ring formation and enhancing loops
are better than the conventional SLR \cite{Gallos15}, 
RBR, and GBR \cite{Park16} methods. 
Especially, in RingLimit5Recal method, 
both high robustness of connectivity and efficiency of paths are 
obtained in saving the resource of links and ports, 
even though the number of additional ports is significantly restricted 
to a constant $5$ from the previous $O(k_{max}) \sim 10^3$ \cite{Hayashi20}. 
Moreover, 
we have found that the reconstructed networks by healing 
can become more robust and efficient 
than the original network before attacks, 
when some extent of damaged links are reusable or compensated 
as the rate $r_{h} \geq 0.5$.

However, it remains an open question 
what structure is the optimally tolerant against further attacks 
in varying the degree distribution after healing.
Even if our prediction comes true, 
it is not yet known 
what approach is more effective and practical for approximately 
maximizing the FVS.
These challenging problems are beyond the discussion of onion-like structure 
under a given degree distribution \cite{Schneider11}\cite{Tanizawa12}. 
In addition, 
it gives an intensive issue how the healing method should be extended 
to interdependent or multilayer networks as networks of networks. 
On the other hand, in application points of view, 
further investigations will be useful for other networks 
if huge computation is available, 
since our obtained some results seem to depend on 
the special topological structure 
such as PowerGrid with a large diameter $D$ (see Table \ref{table_data}).
The development of distributed algorithms within only local information 
is also important for our self-healing methods.

\section*{Acknowledgments}
This research was supported in part by 
JSPS KAKENHI Grant Number JP.17H01729.
\url{https://kaken.nii.ac.jp/ja/grant/KAKENHI-PROJECT-17H01729/}

\end{document}